\documentclass[12pt]{iopart}

\usepackage{iopams}  
\usepackage{url}
\usepackage{graphicx}
\usepackage[utf8]{inputenc}
\DeclareUnicodeCharacter{20AC}{\EUR}
\usepackage{eurosym}
\usepackage{caption}

\begin{document}

\title[A device to safely perform experiments on radioactivity]{A device to safely perform experiments on radioactivity}

\author{Giovanni Organtini}

\address{Sapienza Universit\`a di Roma \& INFN-Sez. di Roma\\Piazzale Aldo Moro 5 - 00185 ROMA (Italy)}
\ead{giovanni.organtini@uniroma1.it}
\vspace{10pt}
\begin{indented}
\item[]July 2018
\end{indented}

\begin{abstract}
This paper describes a cost effective and safe device to perform realistic experiments about the physics of radioactivity in classrooms. It can be used to study both $\alpha-$ and $\beta-$radioactivity as well as $\gamma-$emitters and shows an extremely realistic behaviour. The device, in the form of a Geiger-M\"uller tube, was tested during a public lecture and it deceived many people, even among physicists and teachers.
\end{abstract}

\vspace{2pc}

Published as Giovanni Organtini 2018 {\em Phys. Educ.} {\bf 53} 065018
%\maketitle

\section{Introduction}
Inquiry Based Science Education (IBSE) is nowadays a consolidated practice, though quantitative studies~\cite{ibse} have not found a striking impact of such a methodology on student's content learning and retention. However, as the authors of the cited study pointed out, ''hands-on experiences [$\ldots$] were found to be associated with increased conceptual learning'' and ''there is a clear and consistent trend indicating that instruction within the [IBSE, ed] investigation cycle [$\ldots$] has been associated with improved student content learning''. 

Modern physics is more and more becoming part of the curriculum of secondary school students and is then desirable to design and conduct experiments in this field. Among the various topics in modern physics, a very basic one is radioactivity. Introducing the physics of radioactivity is important for various reasons. The fact that matter can transform into other species of matter was one of the most important results achieved by Enrico Fermi in his theory~\cite{fermi} on the emission of beta rays from nuclei. Though such a possibility is intrinsic in quantum mechanics, before the work of Fermi the community of physicists believed that electrons emitted from nuclei in beta decays pre-existed in the nuclei. Experimenting with beta decays, then, provides a stunning evidence that a physics process can be described in terms of an operator ${\cal O}$ acting on a state $\left|i\right\rangle$, transforming the latter into a different state $\left|f\right\rangle={\cal O}\left|i\right\rangle$. Moreover it provided the first clear evidence that physics processes can be described in terms of intrinsically random variables (not because of our limited knowledge, as in the kinetic theory of gases). The physics of radioactivity proves also to be a good starting point to discuss differential equations by means of the radioactive decay law. 

Unfortunately, to perform experiments in this field one needs to use radioactive sources that, manifestly, can be a problem. Radioactive materials are hazardous and require careful manipulation and conservation. It is also subject to demanding rules for their purchase, conservation and disposal. Detectors for particles emitted in the process are demanding, too, both in terms of cost and in terms of complexity and safety (most of them require high voltages).

It is then not surprising that experiments in the field of radioactivity are not very popular both in schools and in colleges and universities. 

The limitations outlined above lead many authors to develop software tools to simulate such a process. Virtual tools, however, does not leave the same impression to students as a real tool does. For this reason we developed an extremely simple and tangible simulator that can be used almost as a real detector to perform experiments. 

\section{The physics to be simulated}
Our device should imitate the behaviour of either a Geiger-M\"uller tube or of a gamma ray detector. In both cases one expects that particles are emitted isotropically by a radioactive source and that their numbers scales as an inverse square law, i.e.

\begin{equation}
N\left(r, \Delta t\right) \propto \frac{1}{r^2}
\end{equation}
where $N\left(r, \Delta t\right)$ represents the number of particles that can be detected in a time interval $\Delta t$ at distance $r$ from the source.

Radioactive decay occurs as a random process following an exponential decay law, as

\begin{equation}
N\left(t\right) = N(0)\exp{\left( -\frac{t}{\tau}\right)}
\end{equation}
where $\tau$ is called the {\em decay time} of the sample. The process can also be described in terms of the {\em half-life}~$t_{1/2}=\tau\log{2}$.

Different substances have different decay times~\cite{decaytimes} and their possible values span for various orders of magnitude: from minutes to billions of years for natural substances. Depending on that, the amount of decays $\Delta N$ occurring in a given, sufficiently short, time period $\Delta t$, can either appear as constant or as linearly or exponentially decreasing with time.

In an experiment with a real radioactive source we expect to measure a number of events per unit time that increases as $r^{-2}$ and, if the decay time is short enough, decreases as $\exp{\left(-\frac{t}{\tau}\right)}$. Decays happen randomly, hence statistical fluctuations are expected. This, then, should be the behaviour of our device. The distribution of events must be uniformly distributed in short time intervals, while the average number of counts should decrease with time and distance of the sample.

\section{Design of the simulator}
The simulator is in fact just a device that measure distances and time. When an object is put in front of it, pretending to be radioactive, the device reacts according to the laws outlined above.

It consists of an Arduino board~\cite{arduino} used to measure the distance of the {\em source} using an ultrasonic sensor like HC-SR04~\cite{hcsr04}, driving an actuator whose nature depends on the type of device the simulator pretends to be. The Arduino platform consists of a microprocessor that can be easily programmed via a dedicated tool. The programmer is expected to write a function called {\tt setup()}, executed once at the beginning of the program, and a function called {\tt loop()} repeatedly executed indefinitely.

The Arduino sketch of the simulator (the program it runs) is designed such that a {\em decay} occur with a probability $P_0\propto\exp{\left(-\frac{t}{\tau}\right)}$, where $t$ is the time elapsed since the beginning of the run. {\em Decay products} are then detected with a probability $P_1 \propto r^{-2}$, where $r$ is the distance between the sensor and the {\em source}. If a product is detected (it happens with a probability $P_0\times P_1$) the actuator is fired causing an action that depends on the type of detector the simulator is intended to mimic.

\begin{figure}
\begin{center}
\includegraphics{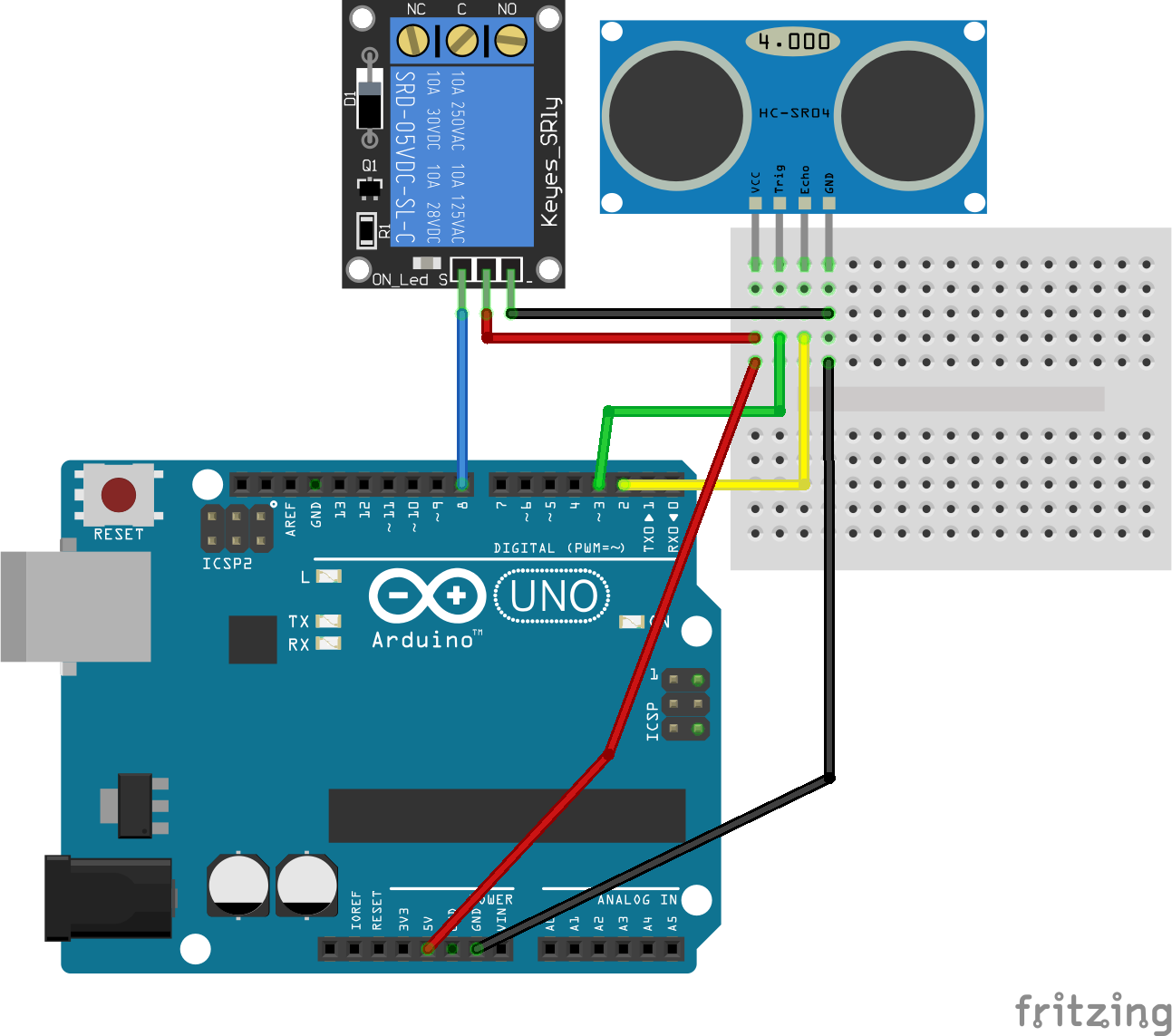}
\caption{\label{fig:drawing}Schema of the simulator.}
\end{center}
\end{figure}
For a Geiger-M\"uller counter, for example, we used just a relay module that, at each detection, switch from open to closed or viceversa. A drawing of the connections is shown in Fig.~\ref{fig:drawing}. The sound produced by the mechanical switch inside the relay module mimic quite realistically the sound of the sparks in a Geiger-M\"uller counter. For a $\gamma$-rays detector, one can mimic the behaviour of a scintillator using an LED that produces a small flash inside a piece of plexiglass. It has been suggested, by people using it, that an LED can be a good choice for people with hearing difficulties. On the other hand, a device based on sound can also be realised using a loudspeaker that has the advantage of not being subject to mechanical ruptures after long operations. Listing~\ref{listing} shows an exert of the Arduino sketch described above and publicly available on GitHub~\cite{sketch}.  

\begin{table}
\begin{verbatim}
void loop() {
  /* measure distance and time */
  float d = measure();
  float t = (millis() - t0)*1.e-3;
  /* compute the probability of a decay */
  float Pdecay = exp(-t/tau);
  float f = (float)random(1000)/1000.;
  /* if an atom decay... */
  if (f < Pdecay) {
    /* ...detect it with a probability that depens on distance d */
    unsigned long trigger = C/(d*d); 
    unsigned long r = random(C);
    if (r < trigger) {
      digitalWrite(CLIK, status);
      if (status == HIGH) {
        status = LOW;
      } else {
        status = HIGH;
      }
    }
  }
}
\end{verbatim}
\caption{\label{listing}Listing of the Arduino {\tt loop()} function. {\tt C} is a constant controlling the effective number of clicks.}
\end{table}
In the listing, {\tt measure()} is a function returning the distance of an object measured with the ultrasonic sensor, while {\tt t0} is assigned at the beginning of the run, in the {\tt setup()} function. We then extract a random number {\tt f} in the interval $[0, 1]$. If such a number is lower than the computed probability decay {\tt Pdecay}, that depends on time, then we evaluate the probability with which the products can be detected. To do so, we extract another random number {\tt r}~$\in [0,1]$ and compare it with {\tt trigger} defined as a constant {\tt C} divided by the measured distance squared. If {\tt r < trigger} we change the status of the relay from open to closed and viceversa. We use the {\tt status} variable to keep track of the current status of the relay.

\begin{figure}
\begin{center}
\includegraphics[width=0.7\textwidth]{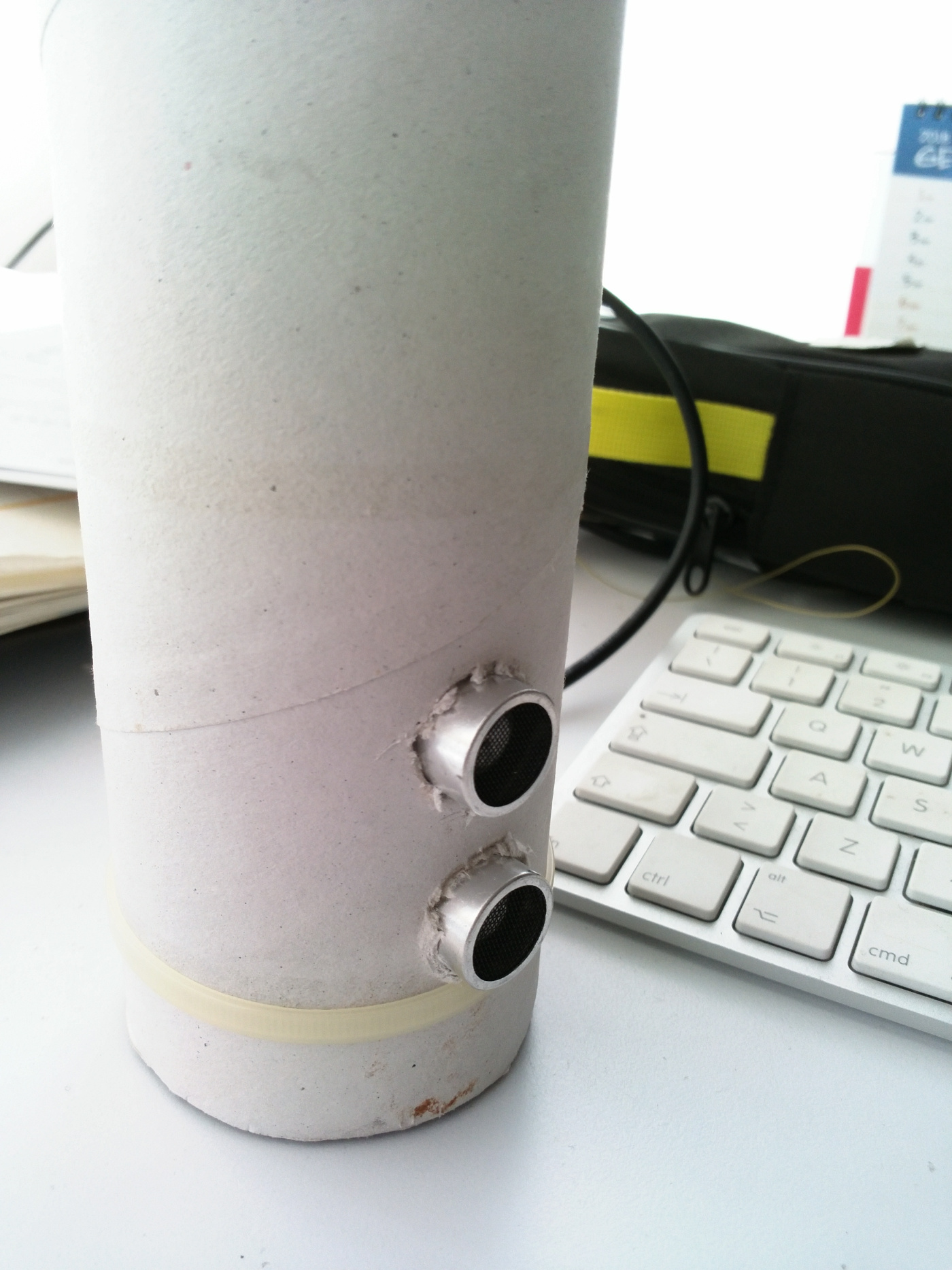}
\caption{\label{fig:simulator}A close up picture of the simulator pretending to be a Geiger-M\"uller tube. The speaker and the microphone of the ultrasonic sensor can be seen protruding from the lateral surface of the cardboard cylinder.}
\end{center}
\end{figure}
All the hardware has been put inside a cardboard cylinder to give it a realistic shape, taking care of leaving the microphone and the speaker of the ultrasonic module to protrude from the lateral surface of the cylinder. Fig.~\ref{fig:simulator} shows the device ready to be used.

The realised device is extremely cost effective: a genuine Arduino board costs as low as 20-25~\EUR{}; an ultrasonic module can be purchased for 3-4~\euro{}, as well as a relay module. The whole project, then, can cost less than 30-35~\euro{}. The total price can even be lowered choosing one of the many Arduino clones.

\section{Results and perspectives}
A short video showing the device in action is available at \url{https://youtu.be/qfedGw0d97g}. The same video is available for download on~\cite{sketch}. The pretended source is in fact a bottle of water on which we put the radioactivity symbol. 

The detector can be used to perform measurements of the count rate versus time and as a function of the distance of the source. In the latter case, one can configure the system such that the decay time $\tau$ is large. That makes the count rate statistically stable with time and it is possible to verify the $r^{-2}$ scaling law. Using shorter decay times introduces a complication consisting in the fact that, while moving the source with respect to the detector, the count rate varies. In order to take into account this effect, it is possible to take a measurement of the decay time keeping the distance between the source and detector constant. Once the decay time is known, it can be taken it into account while measuring the rate as a function of $r$.

\begin{figure}
\begin{center}
\includegraphics{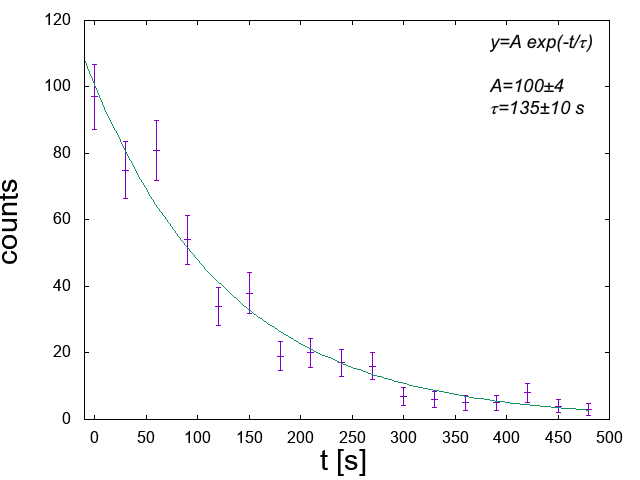}
\caption{\label{fig:decay}The count rate as a function of time in one of the simulator run.}
\end{center}
\end{figure}
Fig.~\ref{fig:decay} shows the data recorded during a run in which we used $\tau=2.24$~min ($^{28}$Al). Counting the number of clicks $\Delta N$ every 30~s we got the data shown in the plot. $\Delta N$ is assumed to fluctuate with a Poisson distribution and error bars are calculated as $\sqrt{\Delta N}$. Such a number can be recorded by hand or other methods like, e.g., recording the sound with a sound recorder and analysing the waveform using software like Audacity~\cite{audacity}. We also provided a tool~\cite{soundcounter} to be used with the {\tt PhyPhoX} App~\cite{phyphox} for smartphones: it counts the number of acoustic events occurred in a given interval of time. 

Data are fitted using an exponential model. The fit has been performed using {\tt gnuplot}~\cite{gnuplot}. The result was $\tau_{exp}=135\pm 10$~s, corresponding to $\tau_{exp}=2.25\pm 0.17$~minutes consistent with the simulator configuration.

The described tool is quite useful when demonstrating the effects of radioactive substances and its cost is low enough to make it affordable to provide a single {\em detector} to each group of few students to let them play with it. The tool can also be used to train students in the statistical analysis of data. They can either take different sets of measurements or regroup those obtained in a single run. In fact it perfectly imitate the behaviour of a real detector providing statistically fluctuating measurements at each run.

Reconfiguring the device at each experiment, choosing different decay times, one can ask students to tell the (pretended) isotopic species of a given sample.

Of course the system can, in principle, be instrumented such that it automatically reconfigure itself, based, e.g., on the detection of some kind of signal coming from the source. In this way, the device can operate exactly as a real detector: it will not react when the obstacle is represented by a non instrumented object, while the decay time can be automatically set recognising the object representing the source. The identification can be made by a variety of methods: bluetooth, RFID, light and so on. Using similar techniques, the device can be instrumented such that one can even simulate the interposition of absorbers between the detector and the source. Such a possibility will be the subject of an improved version of the system.

\section{Summary}
We described a simple and cost effective tool to imitate the behaviour of a radiation detector that perfectly reproduce any aspect of the instrument it pretends to be. The tool, based on the popular Arduino platform, can be used to demonstrate the effects of radioactive substances and to train students in both the field of the physics of radioactivity as well as in statistical analysis of data.

The device is presented in its simplest form, leaving room for interesting improvements.

\end{document}